\documentclass[sigplan,nonacm]{acmart}

\usepackage[english]{babel}
\usepackage{tikz}
\usepackage{amsmath}
\usepackage{xspace}
\usepackage{xcolor}
\usepackage{listings}
\usepackage{multirow}
\usepackage{booktabs}
\usepackage{array} 
\usepackage{makecell}
\usepackage{mdframed}
\usepackage{enumitem}
\usepackage{caption}
\usepackage{subcaption}
\usepackage{array}  

\newcolumntype{C}[1]{>{\centering\arraybackslash}p{#1}}

\usepackage{etoolbox}
\makeatletter
\patchcmd{\maketitle}
	{\@maketitle}
	{\@maketitle\vspace{-4em}}
	{}
	{}
\makeatother
\definecolor{dkgreen}{rgb}{0,0.6,0}
\definecolor{dred}{rgb}{0.545,0,0}
\definecolor{dblue}{rgb}{0,0,0.745}
\definecolor{lgrey}{rgb}{0.9,0.9,0.9}
\definecolor{gray}{rgb}{0.4,0.4,0.4}
\definecolor{darkblue}{rgb}{0.0,0.0,0.6}

\newcommand{\syss}{PITA}
\newcommand{\sys}{PITA\xspace}

\newcommand{\roce}{RoCEv2\xspace}

\newcommand{\signpost}[1]{\textbf{#1.}}

\newcommand{\eat}[1]{}
\newcommand{\todo}[1]{{\color{red}[#1]}}

\newcommand{\ftodo}[1]{}

\newcommand{\tpe}[0]{protocol logic engine\xspace}

\newcommand{\TPE}[0]{Protocol Logic Engine\xspace}
\newcommand{\STPE}[0]{PLE\xspace}

\newcommand{\codify}[1]{{\fontfamily{\ttdefault}\selectfont #1}}

\lstdefinelanguage{cpp}{
  basicstyle=\footnotesize \ttfamily \color{black} \bfseries,   
  breakatwhitespace=false,       
  breaklines=true,               
  captionpos=b,                   
  commentstyle=\color{dkgreen},   
  deletekeywords={...},          
  escapeinside={\%*}{*)},      frame= single,
  framesep = 2pt,
  framexrightmargin=-4pt,
  language=C++,                
  keywordstyle=\color{dblue},  
  morekeywords={event, incoming, app_event, instr_t, addr_t, int8, int32, uint16, uint32, uint8, uint64, list, event_t, event_list, fid_t, extract, set_fid, app_req_t, context, timer_t, dispatch, min, max, len, unseg_data, pkt_event, set_duration, start, mili, add, pkt_bp, checksum16_t, data_t, bit, pkt_t, sliding_wnd, buffer_id_t, data_event, set, first_unset}, 
  identifierstyle=\color{black},
  stringstyle=\color{blue},      
  numbersep=3pt,                  numberstyle=\scriptsize\color{black}\bfseries, 
  xleftmargin=3pt,
  rulecolor=\color{black},        
  showspaces=false,               
  showstringspaces=false,        
  showtabs=false,                
  stepnumber=1,                   
  tabsize=5,      
  belowskip=-15pt,
  aboveskip=10pt,
  title=\lstname, 
}

\begin{document}

\sloppy
\title[]{A Protocol-Independent Transport Architecture}

\author{Kimiya Mohammadtaheri}
\affiliation{%
 \institution{University of Waterloo}
 \country{}
 }

 \author{David Gao}
\affiliation{%
 \institution{University of Waterloo}
 \country{}
 }

 \author{Samuel Zhang}
\affiliation{%
 \institution{University of Waterloo}
 \country{}
 }

  \author{Matthew Chen}
\affiliation{%
 \institution{University of Waterloo}
 \country{}
 }
  \author{Eric Su}
\affiliation{%
 \institution{University of Waterloo}
 \country{}
 }
  \author{Pengyu Ji}
\affiliation{%
 \institution{University of Waterloo}
 \country{}
 }
  \author{Saad Syed}
\affiliation{%
 \institution{University of Waterloo}
 \country{}
 }
  \author{Chris Neely}
\affiliation{%
 \institution{AMD}
 \country{}
 }
  \author{Mario Baldi}
\affiliation{%
 \institution{none}
 \country{}
 }
  \author{Nachiket Kapre}
\affiliation{%
 \institution{University of Waterloo}
 \country{}
 }

   \author{Mina Tahmasbi Arashloo}
\affiliation{%
 \institution{University of Waterloo}
 \country{}
 }

\begin{abstract}
The network transport layer is increasingly implemented in the NIC hardware to meet the performance demands of modern workloads, but this has made it difficult to evolve or deploy new transport protocols. 
Existing approaches either fix protocol logic in the data-path or build protocol-specific assumptions into the architecture that limit the range of protocols that can be supported on a single hardware substrate.

We present \sys, a protocol-independent transport architecture that enables full data-path programmability while sustaining line-rate performance. \sys eliminates protocol-specific assumptions by structuring the data-path around a uniform abstraction over events, state, and instructions, and rethinks core components, including scheduling, packet generation, and data reassembly, to operate on this abstraction.
We evaluate \sys along key dimensions reflecting the goals of its protocol-agnostic datapath design. Specifically, we show that \sys supports diverse protocol semantics by showing it can implement TCP and \roce on the same data path and preserve their distinct end-to-end behavior.
Through targeted microbenchmarks and synthesis on Alveo U250 cards, we show that \syss’s redesigned components sustain high performance under demanding conditions, with modest hardware overhead and meeting timing at 250MHz.
\end{abstract}

\maketitle

\section{Introduction}

The network transport layer sits on the critical path of every networked application. It receives data transfer requests from applications and determines how to reliably and efficiently deliver data over a shared and possibly unreliable network. 
Because it determines the performance achieved by the network and consequently perceived by applications,
the transport layer is repeatedly extended or redesigned as applications, workloads, and network environments evolve. 
Beyond the many TCP variants and optimizations over the years \cite{dctcp, compound, cubic, fast, d2tcp}, the past decade has seen a steady stream of new transport designs, including receiver-driven protocols for data centers \cite{homa, dcpim, phost, ndp}, RDMA over Converged Ethernet (RoCE) \cite{roce}, and transports tailored to emerging workloads such as Amazon's SRD, Google's Falcon, STRACK, and the Ultra Ethernet Consortium's UET \cite{srd, falcon, UET, le2024strack}.

Evolving the transport layer, however, is becoming increasingly difficult. To deliver high throughput ($\ge 100 Gbps$) and low latency (tens of $\mu s$) for modern data center workloads, while minimizing host CPU overhead, key transport functionality is increasingly implemented directly in network interface cards (NICs) \cite{tonic, falcon, srd, UET,limago}.
Such hardware acceleration enables efficient data transfers at line rates of $100Gbps$ and beyond.
However, it also hardens the transport layer: existing hardware transports are either closed-box designs or expose programmable components that are limited in scope or expressiveness. As a result, modifying or deploying new transport functionality often requires modifying low-level hardware designs, making hardware transport far less adaptable than the rapidly evolving workloads it serves.

As such, recent work has explored ways to increase the programmability of hardware transport.
One category exposes programmability \emph{only in the control path} while keeping the protocol logic implemented in the data path fixed (e.g., Falcon \cite{falcon}).
%
Specifically, the data path implements a fixed transport protocol that dictates how per-flow state evolves and how core transport operations such as packet generation, loss detection, recovery, and data reassembly are performed at line rate. 
The control path receives a predefined set of aggregated statistics from the data path and adjusts a fixed set of data-path parameters, such as pacing rates or timeout values.
While the control algorithm can be modified, the protocol logic in the data-path remains fixed.
That is, changing the core transport logic from one protocol to another, for example, evolving RoCE to Falcon, or Falcon to UET, would require modifying the low-level data path hardware design. 

Another category targets data-path programmability.
Unlike the control path, the data path must operate under stringent performance constraints: since it executes stateful transport logic that produces and consumes packets, it must support line rates of hundreds of millions of packets per second.
As such, designing a programmable transport datapath requires carefully balancing flexibility and performance.
Existing systems navigate this tradeoff by incorporating \emph{built-in assumptions} about protocol behavior in their architecture. 
These assumptions constrain which parts of the protocol state are accessible at different decision points, the amount and type of computation that can be performed in response to events, and the kinds of events that can be processed.
While effective for achieving high efficiency, this pushes the transport data path toward one end of the flexibility–performance tradeoff, constraining the range of transport protocols that can be supported on the same hardware substrate (\S\ref{sec:motivation}).



\signpost{Protocol-Independent Transport Architecture} In this paper, we propose a hardware architecture for the network transport layer that enables \emph{full data-path programmability} while sustaining line-rate performance.
Our architecture is inspired by a recent high-level abstraction for network transport protocols \cite{mtp}, which represents the semantics of a transport protocol as mappings from user-defined events and per-flow state to an updated state and a sequence of protocol-agnostic transport instructions.
This abstraction suggests a more uniform structure across transport protocols and motivates revisiting the design of the transport data path to remove protocol-specific assumptions in how events, state, and protocol actions are represented and processed.

Our Protocol-Independent Transport Architecture (\sys) consists of the common components of a transport datapath: event ingestion and scheduling, event processing pipelines, and modules for packet generation, data reassembly, and timer management.
Each component is carefully designed to be either protocol-agnostic with minimal configuration or fully programmable, avoiding partially programmable modules with protocol-specific assumptions, and the interfaces between them follow a protocol-independent abstraction inspired by \cite{mtp}.
Specifically, event ingestion and scheduling, and the packet generation, reassembly, and timer modules are implemented through protocol-agnostic mechanisms that operate on generic events and instructions, and the event processing pipeline is fully programmable to express protocol-specific logic within the same abstraction.

Realizing this design required revisiting each component of the transport data path, as removing protocol-specific assumptions introduces distinct challenges and design considerations across the system.
For example, for event ingestion and scheduling, \sys needs to ingest one incoming event and dispatch one safe-for-processing event per cycle while ensuring state consistency under generic event streams, without relying on any protocol-specific structure or relationships between events or their processing.
\sys addresses this through lightweight coordination between the scheduler and event processors to track event eligibility and enforce state consistency, while minimizing concurrent accesses to the event store buffers and metadata structures.

Similarly, for packet generation, the absence of protocol-specific assumptions requires an instruction-driven design in which instructions fully specify how data is segmented and placed in packets. Sustaining a continuous packet output stream then requires per-instruction data prefetching and efficient interleaving of instructions based on their parameters.
Data reassembly is similarly instruction-driven, managing per-flow reassembly buffers under arbitrary segment placement and according to instruction parameters governing when reassembled data is made available to 
the application.


\signpost{Summary of contributions} 
We revisit the design of hardware transport datapaths to eliminate protocol-specific assumptions while preserving high performance.
In doing so, \sys rethinks both the overall data-path architecture and the design of its individual components,such as event scheduling, packet generation, and reassembly, to operate over generic events, state, and instructions, and shows that a single hardware substrate can efficiently support diverse transport semantics.
Together with prior work on control-path programmability, this paves the way towards a fully programmable network transport layer in the NIC.

\signpost{Evaluation highlights} We evaluate \sys along three dimensions that reflect the key goals of its protocol-agnostic data-path design. First, we show that \sys supports diverse protocol semantics by programming it to implement TCP and \roce, two protocols with radically different semantics, and show that \sys preserves their distinct end-to-end behavior under induced congestion. Second, through targeted microbenchmarks, we demonstrate that the redesigned scheduler, packet generator, and reassembly modules sustain high performance under demanding conditions. Finally, we evaluate \sys's timing and resource utilization on AMD Alveo U250 FPGAs to show that it achieves these capabilities with modest hardware overhead and meets timing at 250MHz.
Prototype will be open-sourced after publication.

\ftodo{TCP and RoCE as examples to show how our architecture supports two very different data paths.}

\ftodo{contribution is revisiting this architecture with the no assumption goal in mind. }

\ftodo{Why not soc-based NIC?}

\ftodo{table for prior work?}

\ftodo{falcon to UET even though they are not too different}
\ftodo{control decisions per-RTT, which is tens of micrseconds?}
\ftodo{deliberately}

\section{Motivating Examples}
\label{sec:motivation}

\ftodo{data path vs control path}

Transport protocols manage reliable and efficient data transfer between endpoints over a shared, possibly unreliable network. Data is divided into segments, identified by their offset within a finite message or a bytestream, and transmitted in individual packets to the receiver. The receiver acknowledges received segments to help the sender track which data has arrived successfully. The sender and receiver cooperate to determine which segments and how many at a time should be (re)transmitted to ensure reliable and fast delivery without overwhelming the network and the receiver. 

Transport protocols have stateful and event-driven data paths.
They keep per-``flow'' state for a large number of flows, where the definition of a flow depends on the protocol, e.g., a connection in TCP, a queue-pair in \roce, or a remote procedure call (RPC) in Homa.
Protocol decisions are triggered by events such as application requests, packet arrivals, and timeouts, and involve updating and analyzing protocol state that summarizes the status of in-flight segments.

To process events at high speed while maintaining state consistency, existing flexible hardware transport data paths incorporate \emph{built-in assumptions} about protocol behavior in their architecture. These assumptions constrain which parts of the protocol state are accessible at different protocol decision points in the data path, the amount and type of computation a protocol can perform when reacting to events, and the kinds of events a protocol can receive.
The following examples illustrate several ways in which these assumptions appear in existing architectures. 

\signpost{Example 1} In Tonic \cite{tonic} and NanoTransport \cite{nano}, the protocol-specific behavior in response to incoming control packets such as ACKs must, respectively, fit within a single pipeline stage (i.e., a single clock cycle, $4ns$ at $100Gbps$) or a feed-forward P4 pipeline whose stages cannot share state and support only a small number of semantically constrained read-modify-write operations.
While this is sufficient for processing simple cumulative ACKs, it cannot support Selective ACKs (SACKs) which are increasingly incorporated into stream-based and message-based protocols to provide richer delivery information and efficiently recover from multiple close-together packet losses \cite{falcon, irn}.
Reacting to SACKs requires multiple dependent state reads, updates, and scans over bitmaps.
Such operations exceed the stateful computation supported by Tonic and NanoTransport. In fact, the Tonic paper and code repository report implementing only a simplified variant that supports a single SACK range using a single find-first-set bitmap operation.

\signpost{Example 2} The data paths in Tonic and NanoTransport are designed around a fixed packet-generation strategy for data packets that constrains how protocol state can be accessed and exposed to the packet-generation logic.
Specifically, both architectures maintain a bitmap in the per-flow state that records only which segments are pending (re)transmission. 
A fixed-function module then uses this bitmap to asynchronously generate data segments, which is followed, in NanoTransport, by a P4 pipeline to manipulate packet headers.
\ftodo{nanotransport additional part}
When reacting to incoming control packets such as ACKs, protocol logic can only convey information to the packet-generation path through this bitmap. The packet-generation logic and subsequent pipelines cannot access certain protocol state, such as receive-side information about which packets have been received so far.
As such, these architectures are not expressive enough for certain common transport behaviors such as piggybacking control metadata about received packets onto outgoing data packets, like RPC completion signals in Homa or ACKs in TCP.
%

\begin{figure}[t!]
\centering
\includegraphics[width=\columnwidth]{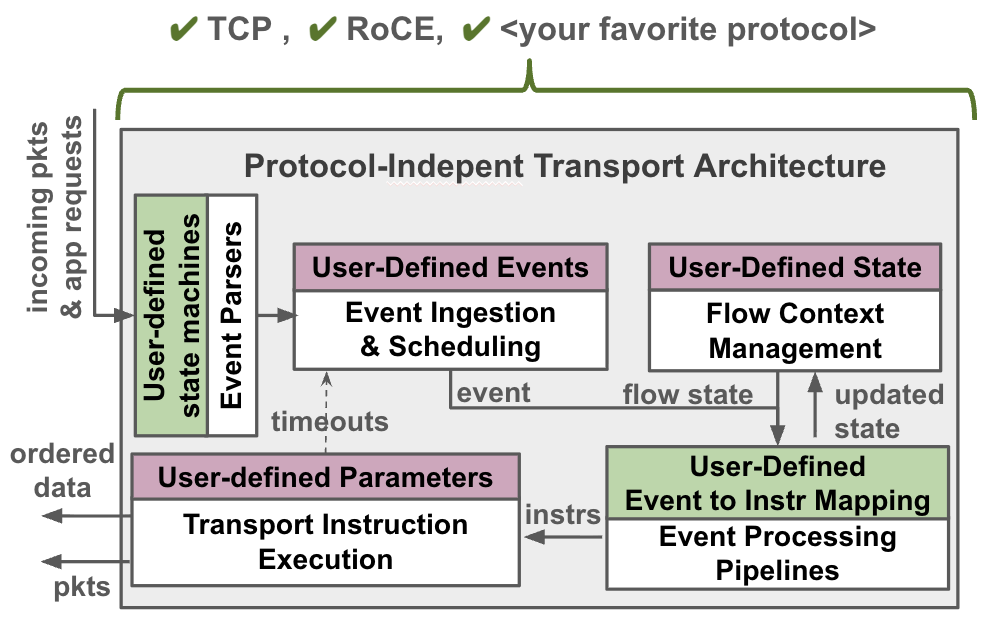}
\caption{\sys's architecture supports protocols with radically different semantics (\S\ref{sec:sys}) (green: fully programmable, purple: protocol-agnostic and reconfigurable)}\label{fig:overview}
\end{figure}

\signpost{Example 3} F4T \cite{F4T} designs a hardware datapath specifically optimized for accelerating TCP.
It observes that, for many TCP variants, incoming event metadata often affects per-flow state in simple ways: metadata values either replace existing state variables or update them using associative operations, such as addition, that can be implemented as a single-cycle read-modify-write.
F4T leverages this observation by accumulating the ``side effects'' of TCP event metadata before invoking the main processing pipeline.
This design implicitly assumes that event effects on protocol state can be summarized using such simple associative updates.
While this assumption holds for many TCP variants, it does not generalize to protocols in which events require more complex state interactions.
For example, merging selective ACKs (SACKs), described above, may require multiple dependent state updates.
Similarly, message-oriented protocols such as \roce or Homa treat each application request as an independent operation with its own parameters and completion state. Unlike TCP send requests, these events cannot be merged and must be tracked individually.

\signpost{Takeaways} Existing flexible hardware transport data paths incorporate built-in assumptions about protocol behavior in their architecture, and as such, restrict the range of transport protocols that can be implemented on the same hardware substrate.

\ftodo{event assumption}

\ftodo{Make sure problems don't look specifically TCP related.}





\section{\sys Overview}
\label{sec:sys}

To provide full data-path programmability without embedding protocol-specific assumptions in the architecture, \sys follows the abstraction introduced in a recent domain-specific language (DSL) for the transport layer \cite{mtp}.
This abstraction represents the semantics of a transport protocol as mappings from an event and the per-flow state to an updated state and a sequence of transport instructions.

Crucially, in this abstraction, events and per-flow state can include arbitrary, user-defined metadata.
Moreover, the transport instructions that perform packet generation, data reassembly, and timer management are designed to be protocol-independent.
They are parametrized to capture the axes along which protocols differ, e.g., header fields and segmentation rules for packet generation, allowing different transport protocols to be expressed using the same instruction set without embedding protocol-specific assumptions.
Finally, the mappings from events and state to instructions are expressed as chains of simple C-like functions with bounded loops and no pointers.
That is, they are only generic constraints that do not impose assumptions about protocol semantics.

\signpost{Protocol-independent architecture} 
By aligning \sys's design with the above abstraction, the same hardware substrate will be able to support a wide range of transport protocols (e.g., TCP and \roce) \emph{without embedding protocol-specific assumptions or logic in the datapath}.
Figure~\ref{fig:overview} provides a high-level view of \sys's architecture.
\ftodo{with reconfigurable modules highlighted in \todo{X} with \todo{X} borders.}  
\sys's modules operate on generic events and per-flow contexts, and a small set of pre-defined, configurable, and protocol-independent transport instructions.
That is, the event-ingestion and scheduling pipeline accepts generic events generated by applications, packets, or timers, while the context table maintains user-defined per-flow state. 
The programmable event-processing pipelines in the \tpe transform events and contexts into updated contexts and transport instructions, and the instruction-execution modules carry out packet generation, reassembly, and timer operations without making protocol-specific assumptions about how or when packets should be generated or reassembled in response to events.

\signpost{Programming \sys} To realize a particular transport protocol on \sys, the user configures a subset of its modules to define how events are parsed, what per-flow context is maintained, and how event-processing pipelines use the user-defined event metadata and context to generate transport instructions. 
We use TCP and \roce as representative examples here (and in our evaluation, \S\ref{eval:e2e}) to demonstrate that \sys can support radically different transport semantics. \sys will naturally accommodate variations within protocol families as well, such as reprogramming features within TCP, RDMA-based, or RPC-based transport.

To implement TCP, the user configures the event parsers in the event ingestion pipeline to extract the metadata required for TCP processing from incoming packets and socket send/receive requests.
The context table is configured with the per-connection state required for TCP processing, such as an integer tracking the first sent but unacked data segment or a bitmap tracking received segments.
The user also configures the timer module with TCP timers to handle lost data or control packets.
The user then programs event-processing pipelines, one for each event.
Each pipeline takes the incoming event and per-flow context as input and produces an updated context together with transport instructions for packet generation, data reassembly, or timer operations.
For example, the user can program the TCP acknowledgment pipeline to compute how many bytes the sender may transmit based on window sizes and the new ACK information, and emit a packet-generation instruction specifying TCP header values, the starting data address, total bytes to send, the maximum segment size, and rules for updating sequence numbers during segmentation.

The user can reconfigure the same architecture to implement a \emph{radically different protocol}, such as \roce.
Unlike TCP, which manages reliable transfer of a byte stream using a sliding window, RoCE supports message-oriented RDMA operations between queue pairs and uses very different mechanisms for loss detection and recovery and congestion control.
Nevertheless, \emph{the same event-processing infrastructure and instruction-execution pipelines} support both protocols.
Specifically, the user can reconfigure the event parsers to now extract metadata from RoCE packets and RDMA work queue elements (WQEs) instead and configure the context table with per–queue-pair state such as message sequence numbers and completion metadata rather than TCP bytestream state.
Similarly, the event-processing pipelines can be reprogrammed to implement RoCE semantics for RDMA read, write, and other operations, emitting the same kinds of instructions but with different parameters, such as packet-generation instructions with appropriate packet and message sequence numbers and payloads, timer instructions to periodically trigger rate 
adjustments, and instructions to reassemble messages and 
generate WQE completion notifications.

Realizing a single hardware substrate that efficiently supports such diverse protocols required revisiting the design of key data-path components, including event scheduling, packet generation, and reassembly. In \S\ref{sec:scheduler} and \S\ref{sec:instructions}, we describe how these components are designed to operate over generic events, state, and instructions. \S\ref{sec:practice} discusses practical considerations, including enforcing atomic execution through backpressure, integrating \sys into a full transport stack with multiple data-paths and per-flow resource management, and opportunities for further optimization.

\ftodo{Add a paragraph that talks about technical insights that we will expand on later in the paper.}

\ftodo{event delivery, state consistency, and instruction realization in Verilog but configurable with parameters. The intention is not for users to fully reprogram these. This is complicated hardware design they should not be exposed to no matter what change. The mapping, fully reporgrammable, HLS is a great fit for.}

\ftodo{holistic view to reprogrammability. Individual components and techniques can be improved over time. a feature of this design}


\ftodo{It has to be crystal clear here what our contribution is over F4T from an architectural perspective, especially since the author in on the PC.}

\ftodo{Right after we introduce the architecture, we should point out that at the high level it is going to look like Falcon or F4T, etc. But their data path is optimized for one protocol, and there is flexibility in the control path. For example, Falcon can change parameters, F4T can change TCP congestion control.}

\ftodo{upgrade each module separately}

\ftodo{Don't want to create expectation of a compiler}

\ftodo{make the broad definition of flow clear. It is kind of buried there now.}

\ftodo{take a pass after technical sections stabilize to ensure accuracy}

\ftodo{Generic?}

\begin{figure}[t!]
\centering
\includegraphics[width=0.9\columnwidth]{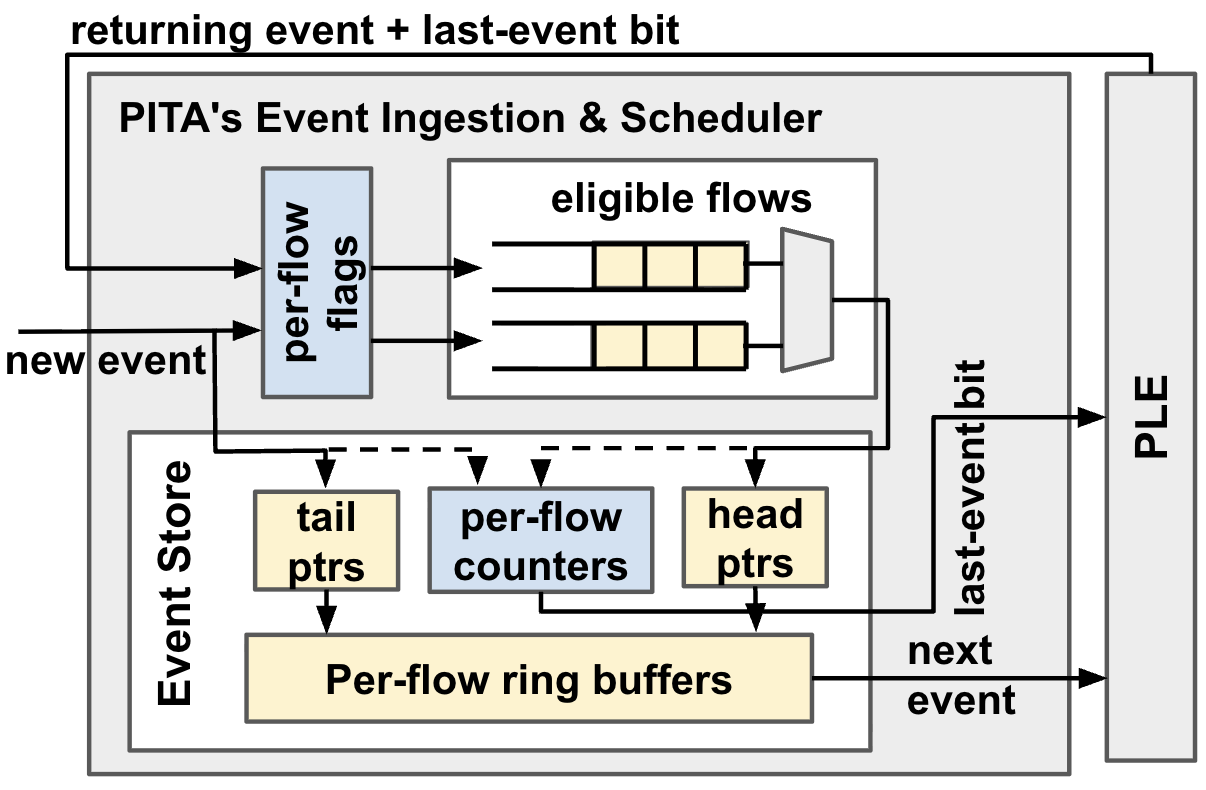}
\caption{\sys's protocol-agnostic event scheduler (\S\ref{sec:scheduler})(yellow: DP RAMs, blue: registers)}\label{fig:scheduler}
\end{figure}

\section{Event Ingestion and Scheduling}
\label{sec:scheduler}

Transport protocols react to events originating from three sources: packets arriving from the network, application requests for data transfer, and timer expirations. 
Among these, packet arrivals are the most frequent and may occur at line rate, while application request rates vary across applications depending on how network-intensive their execution is, and timer events are typically less frequent. 

\ftodo{emphasize lightweight, and that it is necessary to be generic for individual events and high-speed}

To support a wide range of transport protocols without assuming specific event semantics or aggregation behavior, \sys's event-ingestion and scheduling pipeline treats each event as a generic individual entity and is designed to sustain high throughput.
In particular, the pipeline aims to ingest one event per cycle and dispatch one event per cycle to the \tpe whenever at least one eligible event exists.
An event is eligible for processing only if doing so preserves per-flow state consistency. Specifically, events belonging to the same flow must be processed in order, and a new event from a flow cannot enter the processing pipeline while a previous event from that flow is still being processed. 

To enforce these constraints while maintaining high throughput, \sys organizes incoming events into per-flow FIFO queues that are maintained in an \emph{event store} and uses an \emph{event scheduler} that tracks flows with eligible events and selects one each cycle (Figure~\ref{fig:scheduler}).
Implementing this design efficiently requires careful management of memory accesses to the event queues and the metadata structures in the event store and scheduler to minimize concurrent read and write accesses to each stateful module while sustaining one event insertion and one event dispatch per cycle.

\signpost{Efficient tracking of eligible events} 
Because events belonging to the same flow must be processed in order, \sys schedules events at the granularity of flows. A flow is considered \emph{eligible} when (1) its event FIFO is non-empty and (2) no previous event from that flow is currently in the event-processing pipeline. Each cycle, the scheduler selects an eligible flow ID and requests the corresponding event from the event store, which dequeues and returns the event at the head of that flow’s event queue.

Tracking eligibility efficiently requires determining when a flow no longer has an event in the pipeline and whether additional events remain queued for that flow. 
Since each flow can have at most one event in the pipeline at a time, one approach would be to associate each flow with a counter initialized to the pipeline depth and decremented every cycle until the event is guaranteed to have exited. The scheduler could then consult the queue occupancy metadata in the event store to determine whether the flow becomes eligible again. 
However, this requires maintaining unnecessary counter metadata and update logic as well as additional accesses to the event-store metadata structures.

Instead, \sys uses a combination of three lightweight mechanisms: returning events to signal pipeline completion, piggybacking queue-state information on dequeued events, and maintaining small per-flow flags to handle concurrent arrivals. Specifically, each event ``returns'' to the scheduler after completing pipeline processing, implicitly indicating that the flow no longer has an event in flight.
When an event returns, the scheduler must determine whether another event from that flow is waiting in the queue. The event store already maintains per-flow queue occupancy metadata to detect full queues, which is accessed on both event insertion and dequeue. Rather than reading this metadata again, \sys piggybacks this information on the dequeued event by tagging it with a \emph{last-event} bit indicating whether it was the final queued event for that flow.

This information may become stale if new events for the same flow arrive while the event is being processed. As such, \sys maintains two additional per-flow flags that record whether new events have arrived since the flow last became ineligible. Using the last-event bit on the returned event and these flags, the scheduler determines whether the flow should be reinserted into the eligible-flow set.

\signpost{Managing event buffers} 
\sys's event store maintains per-flow event FIFOs using three dual-ported RAMs. One RAM stores the event metadata in a pool of ring buffers (one per flow), while two additional RAMs maintain the head and tail pointers respectively, allowing event insertion and dequeue to proceed independently at line rate.
Specifically, given a ring-buffer index, event insertion reads the current tail pointer, writes the event into the corresponding location in the buffer memory, and updates the tail pointer. Event dequeue follows a similar process using the head pointer.

While queue occupancy could in principle be derived from the head and tail pointers, doing so would require additional pointer arithmetic and break the independence between accesses to the head and tail pointers. Instead, \sys maintains occupancy incrementally using per-flow counters updated on every insertion and dequeue. 
As a result, all updates remain lightweight increments and decrements, and accesses to the head and tail pointers remain independent.

\ftodo{Deallocation strategy can either be something simple, like if a ring buffer is empty for a while, or come from the cache (forward reference to the DRAM section).}

\textbf{Protocol-specific configurations}
tailor the behavior of generic modules to the user's target protocol.
For the event store and the scheduler, which are designed to process generic events, the users only need to provide the maximum expected event width for the target protocol.
%
To extract protocol-specific event metadata, \sys includes a programmable parser placed before the event-ingestion and scheduling pipelines. The parser follows an architecture similar to those used in programmable switches \cite{rmt} and allows users to populate protocol-specific event fields from incoming packets and application requests.

\ftodo{forward reference for data packets}

\section{The \TPE}
\label{sec:tpe}

When an event exits the scheduling pipeline described in \S\ref{sec:scheduler}, \sys retrieves the state associated with the event's flow from a \emph{context table}.
The context table stores per-flow state in a dual-ported RAM. 
Similar to the event-ingestion and scheduling pipelines, the context table remains protocol-agnostic and only needs to know the width of the per-flow state.
\sys then delivers the event and the corresponding per-flow context to the \TPE (\STPE), which applies protocol-specific logic to them.

Conceptually, the \STPE realizes the abstraction described in \S\ref{sec:sys}: it maps an input event and context to an updated context together with a sequence of transport instructions.
The generated instructions are then executed by the instruction-execution modules (\S\ref{sec:instructions}) to do packet generation, data reassembly, or timer operations.
As this mapping differs widely across protocols,
\sys exposes a fully programmable \STPE to the users while keeping the surrounding datapath infrastructure protocol-agnostic and configurable through generic parameters.
Moreover, \sys keeps \STPE's input and output interfaces generic and protocol-independent, so that user-defined protocol logic can seamlessly integrate with the protocol-agnostic scheduler and instruction-execution modules.

In our implementation, the \tpe is programmed using High-Level Synthesis (HLS) \cite{vitis}. Users write a C++ program with a particular structure, and the HLS toolchain generates a pipelined hardware implementation in a hardware description language like Verilog that can be plugged into the rest of the architecture.
In the \STPE program, the user defines a set of event processing functions, one for each event type, that describe how the event interacts with and updates the context of its flow and which instructions should be generated as a result.
To interact properly with \sys's protocol-agnostic modules, the event processing functions follow the interface shown below:

\begin{lstlisting}[language=cpp,
    label=fig:epinterface,
    columns=fullflexible,
    frame=single,
    xleftmargin=0.5em]
typedef hls::stream stream;
void X_event_processor(
    stream<X>&            event_in,   stream<ctx>&          ctx_in, // inputs
    stream<ctx>&          ctx_out,    // updated context
    // instruction streams
    stream<pktGen_instr>  p[N], stream<timer_instr>   t[K],
    stream<reassm_instr>  r[M]);
\end{lstlisting}

All inputs and outputs are expressed using the \codify{hls::stream} type, which the HLS toolchain synthesizes into AXI4-Stream interfaces.

Event and context metadata are delivered to the \tpe as generic bit sequences. 
Users can interpret these bit sequences as protocol-specific metadata by defining appropriate C++ structs.
Similarly, instruction streams use protocol-agnostic metadata structures describing the parameters required for packet generation, reassembly, and timer operations (\S\ref{sec:instructions}).
After processing, the updated context is written back to the context table, and the event, together with the last-event flag, returns to the scheduler to signal that processing for that flow has completed.
\section{Protocol-Agnostic Instruction Execution}
\label{sec:instructions}

This section describes the design of the dedicated modules that execute the transport instructions for packet generation (\S\ref{sec:packet-generation}), data reassembly (\S\ref{sec:reassembly}), and timer operations (\S\ref{sec:reassembly}) that are generated by the \tpe.
To support a wide range of transport protocols without embedding protocol-specific assumptions, these modules operate on generic instruction formats and configurable parameters rather than protocol-specific logic and provide efficient, reusable implementations of common transport operations.

\subsection{Flexible Packet Generation}
\label{sec:packet-generation}

Transport protocols differ in how they construct packets.
Some, such as TCP, assume that all data for a connection resides in a single contiguous bytestream and generate packets by segmenting the available data from that stream into packets carrying byte offsets.
Others, such as \roce, generate packets for a sequence of RDMA operations, where each operation may access different and unrelated memory locations; packet generation must therefore gather data from arbitrary addresses while maintaining a shared packet sequence space across the queue pair.
Protocols such as Homa organize packet generation at the granularity of individual RPCs, with each RPC handled independently rather than as part of a continuous stream or operation sequence.

\signpost{Instruction-driven packet generation}  
In \sys, packet generation makes no assumptions about header formats, data layout in memory, or how packets are derived from larger data units.
Instead, consistent with the abstraction described in \S\ref{sec:sys}, it is driven entirely by the packet generation instructions emitted by the \tpe. Each instruction carries all the information required to generate one or more packets, specifically the location and length of the data to be transmitted, the header to attach, and the parameters needed to update the header across segments, and scheduling and pacing parameters such as per-flow rate or credit.

\begin{figure}[t!]
\centering
\includegraphics[width=\columnwidth]{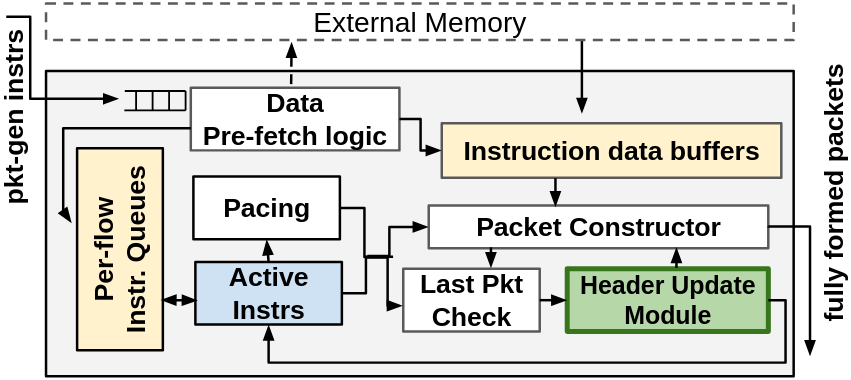}
\caption{\sys's protocol-agnostic packet generation (\S\ref{sec:packet-generation}). (green: reprogrammable, yellow: DP RAMs, blue: registers)}\label{fig:pkt-gen}
\end{figure}

Figure~\ref{fig:pkt-gen} shows a high-level overview of \sys's packet generation module, which organizes execution around per-flow instruction queues, per-instruction data pre-fetch, and interleaved packet construction across flows.
Here, the definition of a flow is configurable to match the protocol's needs. For example, it may correspond to a TCP connection, a \roce queue pair, or an individual RPC message in Homa.
For each flow, the packet generator maintains a queue of instructions and tracks the progress of the currently active instruction.
Moreover, for each instruction, the packet generator maintains a data buffer that is continuously replenished with the payload data from the instruction's specified memory locations.
When an instruction is selected, the packet generator incrementally constructs packets from it according to the instruction parameters. If the instruction requires multiple packets, it remains active across multiple iterations until all data has been transmitted.
To sustain high throughput, the packet generator interleaves instruction execution across flows in a round-robin fashion.

\ftodo{add stage numbers to the descriptions after updating the figure}

\signpost{Continuous payload pre-fetch} 
Data that goes into packet payloads is often stored in external memory. 
\sys avoids stalling on external memory accesses by decoupling data retrieval from packet construction through continuous pre-fetching.
For each instruction, the packet generator maintains a dedicated payload buffer that is populated by a data-fetch engine. When an instruction arrives, data fetching begins immediately to retrieve payload data from the specified memory locations and store it in the buffer. The fetch engine continues to replenish the buffer in the background as packets are generated.
Specifically, further in the pipeline, packet construction incrementally consumes data from this buffer. 
If the buffer's available data falls below a predefined threshold, the data fetch engine is notified for additional fetches, hiding memory access latency and allowing data retrieval and packet construction to proceed concurrently.

\signpost{Instruction arbitration}
The packet generator organizes instructions in per-flow ring buffers implemented using dual-ported RAMs (similar to \S\ref{sec:scheduler}) and schedules packet transmission at flow granularity.
For each flow, \sys maintains the state of the currently active instruction along metadata such as execution progress and optional pacing parameters, specified in the instruction, in registers.
%
%
This information is used by a pacing module to schedule instructions for packet construction.
To avoid head-of-line blocking, an instruction can be preempted after generating a configurable amount of data or packets to return from packet construction, update its execution state, and be rescheduled by the pacing module.
%
%
%
%
%
%

\ftodo{managing instruction queues}

\ftodo{Couldn't assume any relationship between instructions. Had to design without assuming relationship. of course you can further optimize see discussion}

\signpost{Customizable segmentation and packet construction}
The packet constructor receives the parameters of the selected instruction from the scheduler, consumes payload data from the per-instruction buffer, and generates packets according to the specified segmentation parameters.
For each packet, it reads one segment of pre-fetched data (with segment size defined by the instruction), attaches the specified header, and transmits the resulting packet.
If the instruction parameters describe more data than fits in a single packet, the constructor iteratively produces multiple packets, consuming buffered data and triggering additional data fetches as needed to maintain continuous execution.

A key component of this process is a configurable header-update module that determines how header fields evolve across segments.
Rather than assuming a fixed update rule, the module can be configured to apply protocol-specific rules that take in the current header, bytes transmitted so far from this instruction, and instruction parameters, and produce the next header.
For example, in TCP, this component can be configured to add the instruction segment size to the sequence number of the current header to derive the sequence number for the next packet. For RoCE, it can be configured to set opcodes depending on whether the packet is the first, the middle, or the last one in an operation.

\subsection{Data Reassembly and Timers}
\label{sec:reassembly}

Incoming data packets may arrive out of order or with gaps, and transport protocols differ in how they determine ordering and when data is complete and ready for application consumption.
Consistent with the abstraction in \S\ref{sec:sys}, \sys does not make any assumptions about ordering or completion. Instead, the \tpe (\STPE) explicitly specifies these semantics through reassembly instructions.

\signpost{Instruction-driven reassembly} When data packets pass through the programmable parser (\S\ref{sec:scheduler}), their payloads are temporarily stored in a ring buffer in memory in arrival order.
For each packet, the address of its corresponding payload in that memory is provided to the \STPE as part of the event metadata.
Based on protocol-specific logic, the \STPE determines the correct placement of each segment and issues an \emph{add-data-seg} instruction with the address of the segment in the temporary memory and the correct offset as parameters.
The reassembly module then fetches the payload from the temporary memory and inserts it at the specified offset in a per-flow reassembly buffer.
The reassembly module also supports a \emph{flush-and-notify} instruction, which makes a contiguous portion of the buffer available to the application.
In both cases, the module performs only the operations specified in the instruction, without tracking segment state or inferring ordering, leaving all protocol semantics to the \STPE.

\ftodo{figure?}

\signpost{Handling segment alignment}
Executing the \emph{add-data-seg} instruction involves fetching the segment from temporary payload memory and inserting it into the reassembly buffer in a dual-ported RAM at the specified byte offset.
\ftodo{byte-addresssable, limited I/O throughput}
Reassembly buffers are organized in memory-addressable fixed-size chunks. 
Because \sys does not assume any alignment guarantees from instructions, the reassembly module must support arbitrary byte offsets. 
To do that, it aligns incoming data with chunk boundaries using a shift pipeline and performing read-modify-write for the first and last partial chunks when needed.
This process is largely pipelined, where inserting a segment spanning $N$ chunks takes $N+1$ cycles. The extra cycle arises from misalignment and boundary reads, needed to support arbitrary offsets.
Larger chunk sizes require supporting a wider range of shifts; we use 64B chunks, matching the minimum packet size.

For example, consider inserting a 256B segment at offset 71 into a buffer with 64B chunks.
The shift amount is given by the offset modulo the chunk size, i.e., $71 \bmod 64 = 7$.
The segment spans four chunks, which are passed through a shift pipeline that applies shifts based on the set bits in the bit representation of the shift amount (here, shifting by 1, 2, and 4 bytes).
Starting from the second chunk, the pipeline concatenates the previous and current chunks, shifts them together,
and selects the first 64 bytes of the result. This ensures that any leftover data from shifting the previous chunk is correctly accounted for.
For the first and potentially last partial chunks, existing buffer contents must be preserved. In this example, the first 7 bytes of the first chunk and the last 57 bytes of the final chunk are read from memory, merged with the shifted data, and written back.

\ftodo{64B is a good and standard number}

\signpost{Flush and notify}
Once the \tpe determines that a contiguous region of data of size $X$ is ready, it issues an instruction to transfer $X$ more bytes from the reassembly buffer to an application-provided memory address.
To support this, the reassembly module maintains a per-flow read pointer into the buffer and, upon receiving a \emph{flush-and-notify} instruction, outputs the required number of chunks starting from this pointer and advances it so that it always points to the first chunk with data not yet exposed to the application.
This allows the reassembly module to provide the required data for transfer to the application solely based on instruction parameters and without interpreting protocol-specific completion conditions.

\ftodo{64-byte, and 128-byte back-to-back}

\signpost{Timer instructions} Protocols differ in the number of per-flow timers and when they start, restart, and stop them. \sys's design allows users to specify the number of per-flow timers and manage them through instructions issued by the \tpe.
These timers rarely need to operate at a granularity finer than a few microseconds, and their management follows a design similar to prior work \cite{tonic}.

\section{Practical Considerations}
\label{sec:practice}

We have described how \sys achieves protocol-agnostic data-plane programmability by combining generic event scheduling (\S\ref{sec:scheduler}), a programmable \tpe (\S\ref{sec:tpe}), and protocol-independent instruction execution modules (\S\ref{sec:instructions}).
Together, these components provide a flexible substrate for implementing a wide range of transport protocols.
In this section, we discuss additional practical considerations. 

\signpost{Ensuring atomic event processing}
Once an event exits the scheduler, it must be processed atomically. 
That is, if it updates flow state in the \STPE, its generated instructions must not be dropped downstream.
This could happen despite \sys's components being pipelined because one event may generate an instruction whose execution spans multiple cycles.
This multi-cycle execution is inherent to transport data paths: packet generation or data insertion into reassembly buffers involve data movement and are constrained by I/O bandwidth (e.g., transmitting a 1500B packet takes 24 cycles for 64B bus width).
To ensure atomic processing, instruction execution modules provide backpressure signals when their buffers exceed a threshold.
The scheduler then temporarily withholds events from affected flows, maintaining a queue of otherwise eligible flows.
%
These flows are re-enabled once there is less downstream queue buildup, ensuring events are issued only when their instructions can be fully executed.

%

\signpost{Managing per-flow resources}
\sys maintains per-flow resources across components, including scheduler event buffers, context table state, and instruction/data buffers in execution modules.
Flow identifiers in events and instructions are mapped to indices identifying the corresponding per-flow resources in each component, which must be dynamically managed in long-running systems.
The current prototype statically provisions these mappings to focus on demonstrating the programmability of the datapath. However, \sys is compatible with standard dynamic allocation techniques.
For queue-based structures such as event and instruction buffers, a freelist-based allocator can be used to assign buffers to active flows, with buffers entering a draining state upon deallocation and returned to the freelist once empty.
Per-flow state can be swapped to and from off-chip memory (e.g., DRAM) to support larger working sets.
Incorporating these mechanisms into \sys requires modifying some the scheduling and resource management logic.
However, since those rely on protocol-agnostic information such as flow identifiers and generic queue operations, they can be incorporated into \sys without introducing protocol-specific assumptions into the data path.

\signpost{Further optimizations} 
While \sys operates on fine-grained events and instructions, it can support domain-specific optimizations such as event and packet coalescing.
Such coalescing is not universally applicable across protocols (\S\ref{sec:motivation}), and is therefore not embedded in \sys's base architecture.
However, consistent with \cite{mtp} and analogous to segmentation rules, the scheduler and packet generator can expose interfaces for specifying protocol-specific coalescing policies.
Moreover, \S\ref{sec:sys}–\S\ref{sec:instructions} describe a single transport data path. To scale further, multiple \sys instances can be used as parallel datapaths with a load balancer assigning flows to instances \cite{F4T}. Integrating \sys into such systems is an interesting avenue for future work.

\section{Evaluation}

We evaluate \sys along key dimensions that reflect the goals of its protocol-agnostic data-path design: (1) Support for diverse protocol semantics using TCP and \roce as representative examples (\S\ref{eval:e2e}), (2) The ability of the redesigned core components to sustain efficient, line-rate operation (\S\ref{eval:benchmarks}), and (3) Timing and resource overheads in hardware (\S\ref{eval:hw}).

\signpost{Implementation and evaluation setup} \sys's prototype (to be open-sourced) is implemented in 5367 lines of System Verilog code. Users program the \STPE using HLS, and the resulting event pipelines are plugged in with the other core components.
All results are obtained using the AMD Vivado toolchain \cite{vivado}, including its cycle-accurate simulator, which models pipeline and memory behavior at cycle granularity, at a target frequency of 250MHz.

\subsection{Supporting Diverse Transport Semantics}
\label{eval:e2e}

To demonstrate \sys’s ability to support diverse protocol semantics on a single hardware substrate, we program it to implement TCP and \roce (w/ DCQCN). 
These two protocols have \emph{radically different semantics} and together cover a broad \emph{range of common mechanisms} used in transport protocols that shape data-path behavior, including stream- vs. message-oriented semantics, different segmentation and packetization strategies, different kinds of acknowledgements, different pacing mechanisms (window vs. rate) (see \S\ref{sec:sys} for details).

Programming \sys involves configuring the main datapath components and implementing the event processing logic using HLS, as described in \S\ref{sec:sys} and \S\ref{sec:tpe}.
While all components are configured per protocol, we focus on the HLS implementation within the \STPE, which is responsible for mapping user-defined events and flow context to transport instructions and updated state according to protocol logic.
Table~\ref{tab:express} summarizes the results. 
The two protocols differ significantly in the number of events, complexity of the event-processing logic, and pipeline depths, and how they parameterize transport instructions, reflecting their distinct semantics.
For example, \roce requires a larger set of events and more complex processing due to its operation-based model and richer packet semantics, while TCP relies on a smaller set of events but deeper processing pipelines.
Despite these differences, both protocols are implemented within the same protocol-agnostic datapath and incur modest hardware overhead.

\begin{table}[t!]
  \centering
  \normalsize
  \setlength{\tabcolsep}{4pt}
  \begin{tabular}{lll}
    \textbf{} 
      & \shortstack{\textbf{TCP}\\\textbf{(w/ AIMD)}} 
      & \shortstack{\textbf{\roce}\\\textbf{(w/ DCQCN)}} \\
    \midrule
    \textbf{User-defined events} & 4 & 17 \\
    \textbf{HLS LoC} & 612 & 1434 \\
    \textbf{Pipeline Depth (Max)} & 12 & 6 \\
    \textbf{Pipeline Depth (Avg)} & 4.75 & 2.35 \\
    \textbf{FF Usage} & 13384 ($<1\%$) & 132731 ($3.8\%$) \\
    \textbf{LUT Usage} & 7346 ($<1\%$) & 51972 ($3\%$) \\
    \bottomrule
  \end{tabular}
  \caption{TCP and \roce implementation in \sys's \tpe. Each pipeline maps an event and input flow context to a sequence of instructions and the updated context (\S\ref{sec:tpe}). Despite radically different semantics and structure, both are realized within the same protocol-agnostic datapath.}
  \label{tab:express}
\end{table}

\begin{figure}[t!]
\centering
\includegraphics[width=0.78\columnwidth]{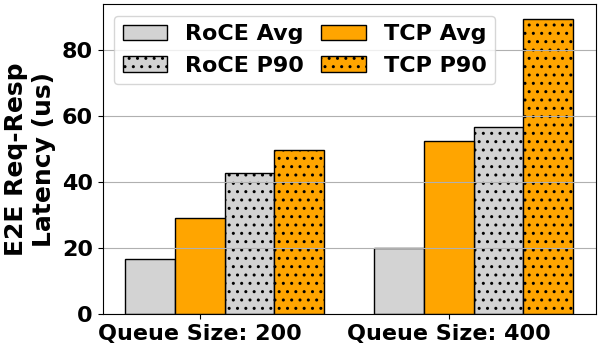}
\caption{Validating faithful realization of TCP and \roce in \sys by comparing their end-to-end behavior for a key-value store application under induced congestion (\S\ref{eval:e2e}).}\label{fig:e2e}
\end{figure}

\begin{figure*}[t]
    \centering
    \begin{subfigure}[t]{0.32\textwidth}
        \centering
        \includegraphics[width=\linewidth]{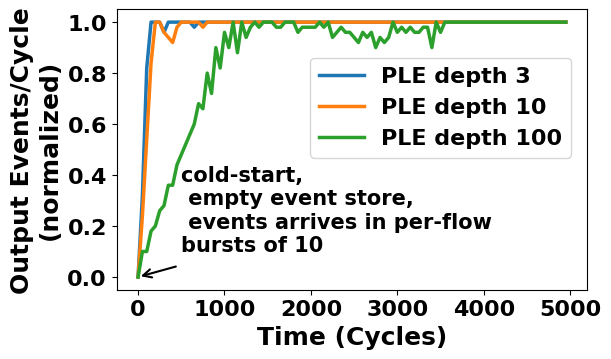}
        \caption{Varying PLE pipeline depth}
        \label{fig:sched-cycle}
    \end{subfigure}
    \hfill
    \begin{subfigure}[t]{0.32\textwidth}
        \centering
        \includegraphics[width=\linewidth]{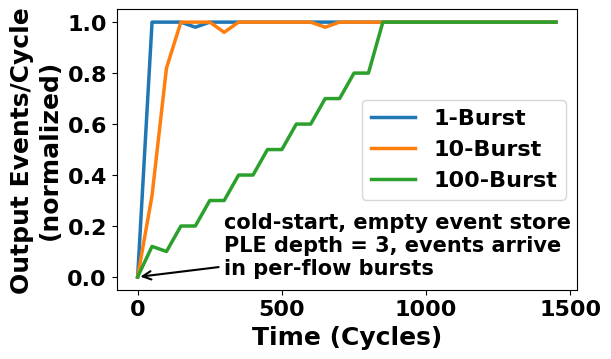}
        \caption{Varying per-flow event burst size}
        \label{fig:sched-burst}
    \end{subfigure}
    \hfill
    \begin{subfigure}[t]{0.32\textwidth}
        \centering
        \includegraphics[width=\linewidth]{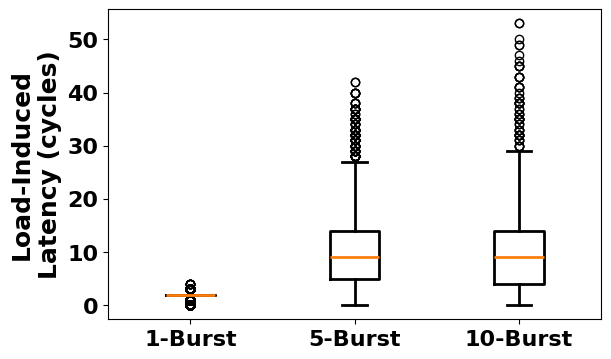}
        \caption{Impact of competing with other flows}
        \label{fig:sched-lat}
    \end{subfigure}
    
    \caption{\sys's protocol-agnostic scheduler sustains line-rate throughput under realistic operating conditions (\S\ref{eval:benchmarks}). While deep \STPE pipelines (a) and extreme burstiness (b) can delay convergence from a \textbf{cold start} and increase intra-flow latency, they do not limit \textbf{steady-state throughput}, and the additional latency due to cross-flow contention (c) remains modest.}
    \label{fig:main}
\end{figure*}

To validate faithful realization of the main mechanisms in these protocols, we evaluate their end-to-end behavior using a simple key-value client-server application.
The client issues back-to-back requests with 4B keys at 8M Req/s, and the server responds with 64B values.
The round-trip time is set to 6.4$\mu$s, and we configure two queue sizes of 12KB and 24KB ($\sim$200 and 400 minimum-sized packets, respectively). For \roce, the ECN marking threshold for DCQCN is set to 3KB ($\sim$ 50 minimum-sized packets).
To induce congestion, we temporarily reduce the queue drain rate to half the request generation rate, causing queue buildup before restoring the original rate.
This high-load, small-message setting combined with induced congestion and the possibility of bufferbloat creates conditions under which the two protocols manifest different latency behavior, allowing us to validate that \sys faithfully captures their semantics.

Figure~\ref{fig:e2e} shows the resulting request–response latency for TCP and \roce under the two buffer sizes.
As expected, \roce achieves lower average and tail latency (p90) under 
congestion, with the gap increasing at the larger queue size. 
This reflects the different transport mechanisms in the two protocols and shows that \sys  preserves protocol-specific behaviors on the same protocol-agnostic hardware substrate.

\ftodo{takeaways?}

\begin{figure*}[t]
    \centering
    \begin{subfigure}[t]{0.32\textwidth}
        \centering
        \includegraphics[width=\linewidth]{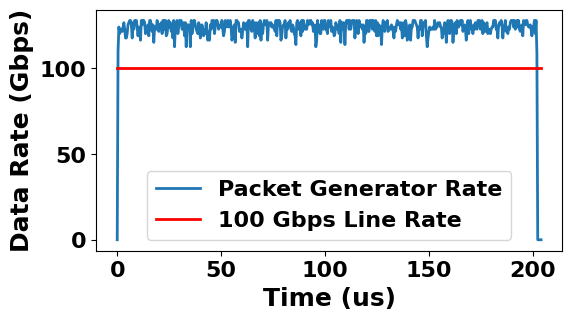}
        \caption{B2B instructions, mixed packet size}
        \label{fig:pktgen-rand}
    \end{subfigure}
    \hfill
    \begin{subfigure}[t]{0.32\textwidth}
        \centering
        \includegraphics[width=\linewidth]{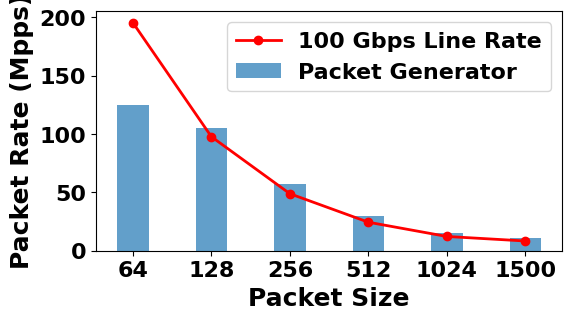}
        \caption{B2B instructions, fixed packet size}
        \label{fig:pktgen-fixed}
    \end{subfigure}
    \hfill
    \begin{subfigure}[t]{0.32\textwidth}
        \centering
        \includegraphics[width=\linewidth]{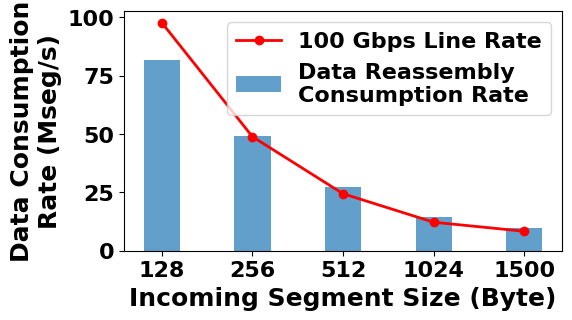}
        \caption{B2B instructions, fixed segment size}
        \label{fig:reassemble}
    \end{subfigure}
    
    \caption{\sys's protocol-agnostic and instruction-driven packet generation and data reassembly sustains line rate under demanding conditions: back-to-back single-packet/segment instructions of various sizes. \S\ref{eval:benchmarks} discusses results and edge cases.}
    \label{fig:main}
\end{figure*}

\subsection{Sustaining Line-Rate in Core Components}
\label{eval:benchmarks}

\ftodo{signpost}

\signpost{Event ingestion and scheduling}
To sustain line rate, \sys's protocol-agnostic event scheduler needs to ingest one event and dispatch one event per cycle to the \STPE if there is at least one eligible event whose processing will not violate state consistency (\S\ref{sec:scheduler}).
Two factors primarily influence the scheduler's event throughput: the \STPE pipeline depth and event arrival patterns.
\STPE pipeline depth impacts how long a flow remains ineligible after issuing an event, and event arrival patterns, particularly the number of active flows and burstiness, affect the availability of eligible events.
These factors are especially important during cold start, when the event store is initially empty and the scheduler must build up enough parallelism across flows to sustain line rate.

\signpost{Impact of \STPE depth}
To evaluate the impact of \STPE depth, we generate events for 1024 flows, where each flow produces bursts of 10 consecutive events, and and measure the scheduler’s output throughput (averaged over 50 cycles) as we vary the \STPE pipeline depth.
Figure~\ref{fig:sched-cycle} shows how throughput approaches line rate from a cold start.
For realistic \STPE depths of 3 and 10 (TCP and \roce have maximum pipeline depths of 12 and 7; Table~\ref{tab:express}), flows become eligible for rescheduling quickly, allowing the scheduler to rapidly approach one event per cycle.
In contrast, with an extreme depth of 100, flows remain ineligible for longer, delaying the buildup of concurrency and increasing the time required to reach line-rate throughput from a cold start.

\signpost{Impact of event burst size} Figure~\ref{fig:sched-burst} shows how the scheduler's throughput approaches line rate from cold start for different per-flow burst sizes, 1024 flows and \STPE depth of 3. 
Burstiness limits the number of flows with eligible events, reducing cross-flow parallelism, especially during cold starts.
For a burst size of one (no burst), the scheduler reaches line rate immediately, because most flows do not have outstanding events in the pipeline (given the number of active flows relative to the \STPE depth), making incoming events immediately eligible for dispatch.
As burst size increases, throughput ramps up more slowly.
This is most evident for a large burst size of 100: early on, only a few flows have events in the scheduler (one in the first 100 cycles, two in the first 200 cycles, and so on).
For more realistic burst sizes (e.g., 10), the scheduler reaches line rate within $\sim$100 cycles.
\ftodo{As such, while burstiness can delay cold-start convergence, the scheduler sustains line-rate throughput once sufficient cross-flow parallelism is available.}

\signpost{Scheduling latency} We use 1024 flows and \STPE depth of 3 to measure how quickly events leave the scheduler after arrival.
As burst size increases, overall scheduling latency increases, as events within a flow must wait for prior events from the same flow to complete processing.
To separate this intra-flow serialization latency from cross-flow contention, we define \emph{load-induced latency} as the additional delay experienced by an event relative to the ``ideal'' scenario in which no other flows are present in the scheduler.
Figure~\ref{fig:sched-lat} shows the distribution of load-induced latency.
We observe that this latency remains relatively small ($\sim$10 cycles on average) and is largely insensitive to burst size. The lower latency observed for burst size one is due to minimal contention, described above, where events are immediately eligible upon arrival.
Reducing intra-flow latency via techniques such as protocol-agnostic but configurable event coalescing is an avenue for future work (\S\ref{sec:practice}) and fits well within \sys's architecture.

\ftodo{tone down?}
\signpost{Event scheduler takeaways} \sys's protocol-agnostic scheduler sustains line-rate throughput under realistic operating conditions.
While extreme burstiness and deep \STPE pipelines can delay convergence from a cold start and increase intra-flow latency, they do not limit steady-state throughput, and the additional latency due to cross-flow contention remains modest. 

\signpost{Packet generation} 
\sys’s packet generation is entirely instruction-driven: each instruction emitted by the \STPE specifies the header, data, and segmentation parameters required to generate one or more packets.
 To evaluate whether the packet generator sustains line-rate operation, we disable rate limiting and generate instructions that each request a single packet with a random size between 64B and 1500B.
 In practice, instructions typically generate multiple packets, and flows are paced by congestion control (\S\ref{eval:e2e}), making this a stress test that exercises the packet generator under more demanding conditions than typical workloads.
 Figure~\ref{fig:pktgen-rand} shows that at a 250MHz clock frequency, the packet generator comfortably sustains 100Gbps line rate under this workload.

Figure~\ref{fig:pktgen-fixed} shows the packet generation rate under a sequence of single-packet instructions with fixed packet sizes 64B to 1500B.
\sys sustains 100Gbps line rate for packet sizes of 128B and above.
For minimum-sized packets (64B), the current prototype does not fully sustain line rate under back-to-back instructions.
This is because transmission of a 64B packet completes in one cycle, requiring the packet constructor to wait for the next instruction for updated header and segmentation parameters. For larger packets, this latency is hidden by packet transmission time.
This gap does not reflect a fundamental limitation and can be addressed using standard techniques such as instruction lookahead or prefetching.
Moreover, this case arises only when flows issue back-to-back instructions, each corresponding to a single minimum-sized packet.
Such patterns are uncommon in well-formed transport protocols, which, instead of issuing back-to-back individual small packets, batch data into multiple full-sized packets.


\signpost{Data reaasembly} 
%
As data packets arrive, the \STPE issues \emph{add-data-seg} instructions that place segments at the correct offsets in per-flow reassembly buffers. To sustain line-rate operation, the reassembly module must therefore consume segments at line rate.
We evaluate this by generating a sequence of \emph{add-data-seg} instructions with fixed segment sizes ranging from 64B to 1500B.
Figure~\ref{fig:reassemble} shows that the design sustains 100Gbps line rate for segment sizes of 256B and above.
For smaller segments (128B and below), the current prototype does not fully sustain line rate under back-to-back instructions. This is due to an additional cycle required when segment offsets are not aligned with the 64B chunk boundaries of the reassembly buffer (\S\ref{sec:reassembly}), which introduces occasional read-modify-write overhead.
%
In practice, workloads rarely consist solely of back-to-back small segments, and mixed traffic allows this overhead to be amortized, enabling the reassembly module to sustain line-rate operation.

\begin{table}[t!]
\centering
\normalsize
\setlength{\tabcolsep}{3pt}
\begin{tabular}{c|ccc|ccc}
\toprule
\multicolumn{1}{c}{} 
& \multicolumn{3}{c}{E = 16}
& \multicolumn{3}{c}{E = 32} \\
\multicolumn{1}{c}{Flows}
& \makecell[l]{LUT \\ {\small (1.7M)}} 
& \makecell[c]{FF \\ {\small (3.5M)}} 
& \makecell[l]{BRAM \\ {\small (2688)}}
& \makecell[l]{LUT \\ {\small (1.7M)}} 
& \makecell[c]{FF \\ {\small (3.5M)}} 
& \makecell[l]{BRAM \\ {\small (2688)}} \\
\midrule
128  & 18K & 29K & 235  & 18K & 29K & 235 \\
256  & 24K & 35K & 312  & 24K & 35K & 343   \\
512  & 36K & 46K & 496  & 38K & 47K & 527 \\
1024 & 57K & 71K & 835    & 60K & 72K & 898  \\
\bottomrule
\end{tabular}
\caption{Resource utilization on an AMD Alveo U250 card  across flow counts and per-flow event buffer depth ($E$). All meet timing at 250MHz.}
\label{tab:synth}
\end{table}

\subsection{Timing and Hardware Resource Utilization}
\label{eval:hw}

We synthesize \sys on an AMD Alveo U250 FPGA to evaluate its hardware cost and timing characteristics for 128-1024 flows and per-flow event buffer size of 16 and 32 (TCP in \STPE, configuration details in Table~\ref{tab:synthparams} in \S\ref{app:params}).
The dominant contributor to resource consumption is BRAM allocated for per-flow event, instruction buffers, and context.
Instruction buffers can remain relatively shallow and are managed using backpressure (\S\ref{sec:practice}).
This is because each packet-generation instruction typically produces multiple large packets, allowing packet transmission to overlap with instruction execution and keep the data-path busy as long as buffers are non-empty.
As a result, the main scalability knobs we vary are the number of flows and the depth of per-flow event buffers.

Table~\ref{tab:synth} summarizes the resource utilization.
All configurations meet timing at 250MHz.
Despite supporting protocol-agnostic execution, \sys incurs modest overhead, using $<$3.5\% of LUTs and $<$2\% of FFs.
BRAM usage ranges from 8\% for 128 flows to $\sim$32\% for 1024.
The higher usage of BRAM compared to other resources reflects the stateful nature of transport processing and is also observed by prior work \cite{tonic, F4T}.
In practice, a single data path instance is expected to handle a smaller number of active 
flows (closer to 128 than 1024), with techniques such as load balancing across multiple data paths used to support larger workloads (\S\ref{sec:practice}, \cite{F4T}).
\ftodo{Overall, these results show that protocol-agnostic execution can be achieved with modest hardware cost while sustaining high clock frequencies.}

%

\ftodo{bring up this is one data path?}


\section{Related Work}

We discussed flexible hardware transport datapaths such as NanoTransport, Tonic, and F4T in \S\ref{sec:motivation}. Here, we situate \sys within the broader categories of related approaches.

\signpost{Protocol-specific hardware transport}
Several works implement transport protocols in FPGAs or ASICs \cite{srd,falcon,UET,limago}.
These designs are highly optimized for a specific protocol and adapting them to a different protocol typically requires substantial hardware modifications and deep implementation knowledge.
In contrast, \sys provides a protocol-agnostic datapath, allowing users to implement different protocols without modifying the underlying hardware.

\signpost{High-level synthesis for transport} 
EasyNet implements a full TCP stack in HLS \cite{easyNet}.
While HLS offers a higher-level programming interface than RTL, EasyNet's design remains tailored to TCP.
Meeting the stringent performance requirements of transport datapaths requires writing HLS code with hardware constraints in mind (e.g., pipelining and memory layout), making it significantly different from conventional C++ code.
As a result, extending these designs to support other protocols remains challenging and requires substantial re-engineering.


\signpost{SoC-based transport offloads}
Some works offload parts of the transport stack to embedded processors in SoC-based SmartNICs \cite{acceltcp,flextoe}, with the remaining functionality running on the host CPU.
These designs offer a more software-like development model than FPGA or ASIC approaches.
However, they typically cannot achieve the same line-rate performance as FPGA- or ASIC-based designs \cite{accelnet,F4T}.

%

\signpost{Software transport programmability}
Prior work has proposed transport-layer abstractions, primarily in software. CCP \cite{ccp} focuses on congestion control, while MTP \cite{mtp} provides a general abstraction for transport protocols.
\sys builds on these abstractions and shows how they can inform the design of a fully programmable hardware datapath.



\ftodo{Why not do SoC-based NIC?}
\section{Conclusion}

We present \sys, a protocol-independent transport architecture that removes protocol-specific assumptions from the data path while sustaining high performance. By structuring the data path around a uniform abstraction over events, state, and instructions, \sys demonstrates that flexibility and efficiency need not be at odds in hardware transport design.
We believe this approach opens a path toward fully programmable transport layers in NICs, enabling rapid 
evolution of transport protocols to meet emerging workloads.

\bibliographystyle{ACM-Reference-Format}
\bibliography{reference}

\appendix
\section{\sys Parameters}
\label{app:params}

Table~\ref{tab:synthparams} shows the relevant \sys parameters.

\begin{table}[h!]
\centering
\small
\setlength{\tabcolsep}{3pt}
\begin{tabular}{l l l}
\toprule
\textbf{Module} & \textbf{Parameter} & \textbf{Value} \\
\midrule

\multirow{6}{*}{Global}
& flow count & ED \\
& event width & 64 b (TCP) \\
& event type count & 4 (TCP) \\
& context width & 938 b (TCP) \\
& serialized data width & 64B \\
& serialized packet width & 64B \\
\midrule

\multirow{2}{*}{Scheduler}
& per-flow event buffer depth & ED \\
& eligible flows queue depth & = flow count \\
\midrule

\multirow{3}{*}{Pkt Gen}
& header width & 168 b \\
& per-flow instr. queue depth & 8 \\
& per-flow pre-fetch buffer len. & 64 $\times$ 64B\\
\midrule

Reassembly
& per-buffer reassembly buffer len & 256 $\times$ 64B \\
\bottomrule
\end{tabular}
\caption{Parameter values for results in \S\ref{eval:hw}. ED stands for experiment-dependent}
\label{tab:synthparams}
\end{table}

\end{document}